\documentclass{article}
\usepackage{wrapfig}
\usepackage{amssymb,latexsym,amsmath,color,mathrsfs,ifsym,graphics,stmaryrd}
\usepackage{amsthm, amsfonts, mathrsfs}
\usepackage[colorlinks,linkcolor=blue,urlcolor=blue,citecolor=black,
plainpages=false,pdfpagelabels,breaklinks]{hyperref}
\usepackage[all]{xy}

\usepackage{graphicx}
\usepackage[margin=2.5 cm]{geometry}
\usepackage{caption}

\DeclareMathOperator{\EA}{EA}

\begin{document}

\title{{\sc Tensorial Quantum Mechanics:\\Back to Heisenberg and Beyond}}

\author{{\sc Christian de Ronde}$^{1,2,3}$, {\sc Raimundo Fern\'andez Mouj\'an}$^{2,6}$, {\sc C\'esar Massri}$^{4,5}$}
\date{}

\maketitle
\begin{center}
\begin{small}
1. Philosophy Institute Dr. A. Korn, University of Buenos Aires - CONICET\\
2. Center Leo Apostel for Interdisciplinary Studies\\Foundations of the Exact Sciences - Vrije Universiteit Brussel\\
3. Institute of Engineering - National University Arturo Jauretche\\
4. Institute of Mathematical Investigations Luis A. Santal\'o, UBA - CONICET\\
5. University CAECE\\
6. Philosophy Institute, Diego Portales University, Santiago de Chile
\end{small}
\end{center}

\begin{abstract}
\noindent In this work we discuss the establishment of Standard Quantum Mechanics (SQM) developed through Schr\"odinger's and Dirac's wave-vectorial reformulations of Heisenberg's original matrix mechanics. We will argue that while Heisenberg's approach was consistently developed ---taking as a standpoint the intensive patterns that were observed in the lab--- as an invariant-operational formalism, Dirac's axiomatic re-formulation was, instead, developed ---taking as a standpoint Schr\"odinger's wave mechanics and the methodological guide of Bohr and logical positivists--- as an essentially inconsistent ``recipe'' intended (but unable) to predict (binary) measurement outcomes. Leaving SQM behind and attempting to restore the consistent and coherent account of a real state of affairs, we will present a new tensorial proposal which ---taking as a standpoint Heisenberg' original approach--- will prove capable not only to extend the matrix formalism to a tensorial representation but also to account for new experimental phenomena. 
\end{abstract}
\begin{small}

\end{small}

\newtheorem{theorem}{Theorem}[section]
\newtheorem{definition}[theorem]{Definition}
\newtheorem{lemma}[theorem]{Lemma}
\newtheorem{proposition}[theorem]{Proposition}
\newtheorem{corollary}[theorem]{Corollary}
\newtheorem{remark}[theorem]{Remark}
\newtheorem{example}[theorem]{Example}
\newtheorem{notation}[theorem]{Notation}

\bigskip

\bigskip

\bigskip

\bigskip

\bigskip

\bigskip

\section*{Introduction}

Today there is a general consensus within the contemporary physics community on what is to be considered the ``Standard'' version of Quantum Mechanics (SQM). This is, in fact, what students are taught in Universities all around the world when following a course about QM. When physicists talk about QM they actually have in mind either Schr\"odinger's wave equation from 1926 (with Born's probabilistic interpretation) or Paul Dirac's vectorial re-formulation (heavily influenced by Bohrian and positivist ideas) presented in 1930 and rendered ``more rigorous'' in mathematical terms by John von Neumann in 1932.  Of course, there are many different textbooks with small variations and slightly different presentations,\footnote{It is important to remark that most physicists are not interested at all in the many ``interpretations'' which are heatedly debated in philosophical journals. As Maximilian Schlosshauer \cite[p. 59]{Schlosshauer11} has recently described: ``It is no secret that a shut-up-and-calculate mentality pervades classrooms everywhere. How many physics students will ever hear their professor mention that there's such a queer thing as different interpretations of the very theory they're learning about? I have no representative data to answer this question, but I suspect the percentage of such students would hardly exceed the single-digit range.''} but the core of what is taught by physics Professors is always the same, a set of 8 or 9 postulates, that are common to all textbooks, which attempt to show how these axioms are capable to predict measurement outcomes. As it is well known, within this ``standard'' presentation, it is Heisenberg's matrix mechanics which has been mostly left aside.  

In this work we attempt to critically consider the way in which SQM was established by imposing a series of controversial operations (originated in great part in Bohr's work) and argue not only on the need to return to Heisenberg's original formulation, but also to go beyond it and advance in a tensorial formulation. The paper is organized as follows. In section 1 we present the main ideas which allowed Heisenberg to develop QM. Section 2 addresses the replacement of  matrix mechanics by Schrödinger's differential wave equation. In section 3, we review the vectorial axiomatic re-formulation ---already implicit in Schr\"odinger's proposal--- developed by Dirac in 1930. In section 4 we argue for the need not only to return to Heisenberg's matrix mechanics, but also to extend the mathematical formalism in tensorial terms. In section 5 we present Tensorial QM. In the last section we present some final remarks.

\section{Heisenberg's Matrix Quantum Mechanics}

After Max Planck postulated the discreteness of the {\it quantum of action} during the first year of the 20th Century, it took 25 more years for physicists to reach a closed mathematical formalism that would allow them to account in a quantitative fashion for the intensive line-spectra that were observed in the lab. It was Werner Heisenberg, a young physicist from Munich, who was finally able to create a consistent mathematical formalism capable to capture all observed quantum phenomena in an operationally-invariant manner \cite{Heis25}. In this respect, one cannot stress enough the essential subversive role played by the Machian positivist critique of the {\it a priori} concepts of classical mechanics within Heisenberg's work ---also essential within Einstein's development of realitivity. It was this attitude that allowed him to approach the issue less burdened by the presuppositions derived from classical physics. As described by Sheila Jones: 
\begin{quotation}
\noindent {\small ``Heisenberg was groping in the dark, taking a stab at a new approach, just as all the physicists working on the puzzle of the quantum had been doing for years. He didn't really know what he was doing. He hadn't worked out a firm foundation yet, but he was seeking to describe the behavior of the quantized atom using only energy, intensity and frequency, and the observable experimental results from spectroscopy that showed up in the spectral signature of heated gases.'' \cite{Jones08}}
\end{quotation}
In the previous years Heisenberg had followed the lead of Niels Bohr, focusing on the problem (created by Bohr's model) of describing the trajectories of electrons inside the atom. Bohr, let us recall, trying to stick to the atomist representation familiar to physicists while incorporating the discreteness of the quantum of action, had related his essentially inconsistent yet effective algorithmic model ---capable to predict the spectral lines of a specific experiment--- to the image of a ``small planetary system'' with microscopic particles traveling along discrete quantum orbits around a central nucleus. According to the Danish physicist, electrons would change orbits through an essentially irrepresentable process, a ``quantum jump''. Even though none of this made much sense, nor could it be demonstrated or really explained, physicists were immediately captured by the predictive capacity of Bohr's model, they bought the whole package, and focused on trying to describe the real trajectories of these surreal quantum corpuscles. Heisenberg followed the lead of Bohr for some time, but the critical reaction of Arnold Sommerfeld and Wolfgang Pauli to Bohr's ideas convinced him in 1925 that this path lead to a dead end and that he should forget about the attempt to describe the trajectories of particles (see for a detailed discussion \cite{BokulichBokulich20}). Following the same Machian principle which had helped Einstein to develop special relativity, Heisenberg would reframe the problem completely ---escaping the classical images proposed by Bohr--- focusing exclusively on the intensive quantities that were actually described by experimentalists. As explained by Jaan Hilgevoord and Joos Uffink: 
\begin{quotation}
\noindent {\small ``His leading idea was that only those quantities that are in principle observable should play a role in the theory, and that all attempts to form a picture of what goes on inside the atom should be avoided. In atomic physics the observational data were obtained from spectroscopy and associated with atomic transitions. Thus, Heisenberg was led to consider the `transition quantities' as the basic ingredients of the theory.'' \cite{HilgevoordUffink01}}
\end{quotation}
That same year, emancipating himself completely from classical ideas, Heisenberg \cite{Heis25} would present his groundbreaking results in the following manner: ``In this paper an attempt will be made to obtain bases for a quantum-theoretical mechanics based exclusively on relations between quantities observable in principle.'' Almost half a century later he would narrate how these events had actually unfolded:  
\begin{quotation}
\noindent {\small ``In the summer term of 1925, when I resumed my research work at the University of G\"ottingen ---since July 1924 I had been {\it Privatdozent} at that university--- I made a first attempt to guess what formulae would enable one to express the line intensities of the hydrogen spectrum, using more or less the same methods that had proved so fruitful in my work with Kramers in Copenhagen. This attempt lead me to a dead end ---I found myself in an impenetrable morass of complicated mathematical equations, with no way out. But the work helped to convince me of one thing: that one ought to ignore the problem of electron orbits inside the atom, and treat the frequencies and amplitudes associated with the line intensities as perfectly good substitutes. In any case, these magnitudes could be observed directly, and as my friend Otto had pointed out when expounding on Einstein's theory during our bicycle tour round Lake Walchensee, physicists must consider none but observable magnitudes when trying to solve the atomic puzzle.'' \cite[p. 60]{Heis71}}
\end{quotation}
Max Born \cite[p. 160]{Jones08} would also recall his surprise when his young student gave him a paper to publish in {\it Zeitschrift f\"ur Physik} which included a new non-commutative multiplication rule: ``Heisenberg's rule of multiplication left me no peace, and after a week of intensive thought and trial, I suddenly remembered an algebraic theory that I had learned from my teacher... in Breslau. Such quadratic arrays are quite familiar to mathematicians and are called matrices, in association with a definite rule of multiplication.'' As Heisenberg would later explain: 
\begin{quotation}
\noindent {\small ``The equations of motion in Newtonian mechanics were replaced by similar equations between matrices; it was a strange experience to find that many of the old results of Newtonian mechanics, like conservation of energy, etc., could be derived also in the new scheme. Later the investigations of Born, Jordan and Dirac showed that the matrices representing position and momentum of the electron do not commute. This latter fact demonstrated clearly the essential difference between quantum mechanics and classical mechanics.''  \cite[p. 4]{Heis58}}
\end{quotation}
Heisenberg's theory, which had been developed exclusively from the tables of intensive patterns observed in the lab, was operationally grounded right from the start. Heisenberg was simply trying to provide an invariant mathematical scheme capable to account for the empirical tables of intensive data. Was there any consistency, any invariance to be found between these tables of (intensive) numbers collected through experimental procedures? The answer provided by Heisenberg in terms of a new mathematical scheme was of course a positive one. What is truly remarkable is that in finding this invariance of intensities Heisenberg had rediscovered the theory of matrices. That same year, with the help of Born and Pascual Jordan, Quantum Mechanics (QM) would fianlly find its final form \cite{BornJordan25, BornHeisenbergJordan26}. Let us stop for a minute just to note some aspects of Heisenberg’s proposal. Firstly, that its success was the direct consequence of abandoning, at least for a moment, the Bohrian approach of trying to describe the trajectories of presupposed corpuscles (a path he had attempted before but had lead him nowhere). Secondly, that his formalism was in fact invariant, something which permitted to discuss consistently the operational results found in the lab also when considering different reference frames or bases. Thirdly, that the quantities this formalism was pointing as its invariant elements were intensities (and not binary values); something which, of course, if taken seriously, imposed the need of a radical innovation: to develop a new originally intensive physical concept. And finally, that this formulation implied a mathematical theory to which physicists were not accustomed. In fact, the introduction of a new mathematical theory implied of course a radical shift in physics which, since classical mechanics, had been constrained to solving differential equations. Furthermore, the loss of a continuous representation implied for classically trained physicists a supposedly ``higher level of abstraction'' contrary to the ``intuitive representation'' provided by {\it infinitesimal calculus}. Physicists had no obvious picture they could apply in order to make sense of this new mathematical representation. As a consequence, they would observe the new theory with unease and discomfort. Matrix mechanics was accepted not without resentments to be the adequate theory for quantum phenomena. However, the immediate reaction was to interpret this theory, again, in atomistic terms, although, as Sin Itiro Tomonaga \cite[p. 223]{tomonaga} reports, as they did that, they still had many questions they could not answer: ``[i]f the location of an electron and of other particles were to become such an abstract aggregate, or matrix, how can we explain in this theory the track of a particle observed commonly in a Wilson chamber? It was Lorentz who said: `Can you imagine me to be nothing but a matrix? It is hardly to believe that all this is real'.'' Of course, in retrospect, it should have been clear not only that the atomist representation was not suitable for this theory of intensities, but also that the picture provided by differential calculus and Newton's theory had also been, at the time of its creation ---in contraposition to Aristotelian physics---, exactly the opposite of an ``intuitive picture''. Physicists had simply become accustomed to the classical representation which they now regarded as ``commonsensical'' or ``self-evident''. So even though Heisenberg's formalism was capable of describing in invariant terms the intensive values observed in the lab, this did not seem to be enough. Physicists were simply not prepared to give up on the modern atomist spatiotemporal representation, even if the cost was as high as to accept the existence of Bohr's strange and irrepresentable ``quantum particles'' and ``jumps''. It is in this context that Schr\"{o}dinger's new differential formulation of wave mechanics would be received by the physics community with enormous relief.

\section{Schr\"odinger's ``Wave Like'' Differential Reformulation}

Heisenberg's paper was sent in July 1925 and published in December that same year. But just one month later, in January 1926, Erwin Schr\"odinger would present ---following the work of Louis de Broglie--- a new ``wave like'' differential equation which was also able to compute the correct energies for the quantized hydrogen atom. Schr\"odinger would complete his formulation in a series of four papers published in {\it Annalen der Physik}, later on collected in a single work \cite{Schr28}. Of course, the choice of the physics community between Heisenberg's ``weird'' matrices and Schr\"odinger's good old comfortable differential equation was not really difficult. While Schr\"odinger's proposal promised to restore a continuous space-time representation within a well known mathematical scheme, Heisenberg's calculus seemed to lack any picture that would explain what was really going on with quanta, reinforcing in turn the unacceptable elements of discontinuity. In words of Max Jammer: 
\begin{quotation}
\noindent {\small ``After the conceptual cataclysm evoked by the latter [matrix mechanics] it seemed as if Schr\"{o}dinger's return to quasi-classical conceptions reinstated continuity. Those who in their yearning of continuity hated to renounce the classical maxim \emph{natura non facit saltus} acclaimed Schr\"{o}dinger as the herald of the new dawn. In fact, within a few brief months Schr\"{o}dinger's theory `captivated the world of physics' because it seemed to promise `a fulfillment of that long-baffled and insuppressible desire. Einstein was `enthusiastic' about it, Planck reportedly declared `I am reading it as a child reads a puzzle', and Sommerfeld was exultant.''  \cite[p. 269]{Jammer89}} 
\end{quotation}

However, this conservative attempt to restore a somewhat classical representation of physical reality within the theory of quanta would immediately find evident limits for its consistent development. Schr\"odinger himself would recognize the difficulties: the domain of the wave equation he had constructed was not 3-dimensional but {\it configuration space}, and thus the possibility to regain a classical and intuitive spatial representation became far from obvious.   
\begin{quotation}\noindent {\small ``The true mechanical process is realized or represented in a fitting way
by the \emph{wave process} in \emph{q}-space, and not by the motion
of \emph{image points}  in this space. [...] In \emph{this} sense do
I interpret the `phase waves' which, according to de Broglie,
accompany the path of the electron; in the sense, therefore, that no
special meaning is to be attached to the electronic path itself (at
any rate, in the interior of the atom), and still less to the
position of the electron on its path. And in this sense I explain
the conviction, increasingly evident to-day, \emph{firstly},   that
real meaning has to be denied to the phase of electronic motions in
the atom; \emph{secondly}, that we can never assert that the
electron at a definite instant is to be found on \emph{any definite
one} of the quantum paths, specialized by the quantum conditions;
and \emph{thirdly}, that the true laws of quantum mechanics do not
consist of definite rules for the \emph{single path}, but that in
these laws the elements of the whole manifold of paths of a system
are bound together by equations, so that apparently certain
reciprocal action exists between the different paths. [...] All
these assertions systematically contribute to the relinquishing of
the ideas of `place of the electron' and `path of the electron'. If
these are not given up, contradictions remain. This contradiction
has been so strongly felt that it has even been doubted whether what
goes on in the atom could ever be described within the scheme of
space and time.''  \cite[p. 25-26]{Schr28} (his emphasis).}
\end{quotation}
Adding to the already complex situation, very soon the battle between wave and matrix mechanics would face an essential difficulty for both parties. After his second paper, and for the surprise of everyone in the physical community, Schr\"{o}dinger demonstrated what he called ``a formal, mathematical identity'' between Heisenberg's matrix mechanics and his own wave mechanics \cite{Sch5}. In the introduction he remarked that both approaches seemed in principle fundamentally different in mathematical methods, assumptions and presentation: 
\begin{quotation}\noindent {\small
``Above all, however, the departure from classical mechanics in the two theories seem to occur in diametrically opposed directions. In Heisenberg's work the classical continuous variables are replaced by systems of discrete numerical quantities (matrices), which depend on a pair of integral indices, and are defined by \emph{algebraic} equations. The authors themselves describe the theory as a `true theory of a {\it discontinuum}'. On the other hand, wave mechanics shows just the reverse tendency; it is a step from classical point-mechanics towards a \emph{continuum-theory}. In place of a process described in terms of a finite number of dependent variables occurring in a finite number of total differential equations, we have a continuous \emph{field-like} process in configuration space, which is governed by a single \emph{partial} differential equation derived from a principle of action.'' \cite{Sch5}}
\end{quotation}
What is important to notice, for our purposes, is that Schr\"{o}dinger had only demonstrated that his wave mechanics implied the matrix formulation, but the converse was still missing. The reason, as shown in \cite{deRondeMassri24}, was that wave mechanics was, in fact, included within Heisenberg's matrix formulation. While Heisenberg's mathematical theory referred to all matrices, of any rank, Schr\"odinger's formulation restricted Heisenberg's original operational space only to matrices of rank = 1. Unfortunately, adding to this confusion, Wolfgang Pauli, Carl Eckart and John von Neumann would give their own independent ``proofs of the equivalence'' giving rise to what Fred Muller has correctly termed ``the equivalence myth'' \cite{Muller97}. Useless to say, Heisenberg was not delighted with the equivalence and neither was Schr\"{o}dinger who declared: ``I was discouraged, if not repelled, by what appeared to me a rather difficult method of transcendental algebra, defying any visualization''. 

Short after the acceptance of the establishment of this equivalence between matrix and wave mechanics, Born began to study the scattering of particles by a spherically symmetric potential, developing what is now known as the `Born approximation'. He preferred to make the calculations by means of Schr\"{o}dinger's formulation due to its simplicity for this specific problem. To model the collision, Born proposed ``to solve Schr\"{o}dinger wave equation for the system-plus-atom subject to the boundary condition that the solution in a preselected direction of electron space goes over asymptotically into a plane wave with exactly this direction of propagation (the arriving electron). In a thus selected solution we are further interested principally in a behavior of the `scattered' wave at infinity, for it describes the behavior of the system after the collision''. He solved the problem, showing that ``the perturbation, analyzed at infinity, can be regarded as a superposition of solutions of the unperturbed problem'' and proposed the reading that each coefficient ``$\Phi_{n,m}(\alpha, \ \beta, \ \gamma)$ gives the probability for the electron, arriving from the $z$-direction, to be thrown out into the direction designated by the angles $\alpha, \ \beta, \ \gamma$ [...]''. Born corrected in the same paper, in a note added in proof, that the probability was proportional to the square of the coefficient instead to the coefficient itself. Ever since, this has been known as the \emph{Born Rule} and his reading of the wave function as the \emph{probabilistic interpretation}.
\begin{quotation}
\noindent\small{``Schr\"{o}dinger's quantum mechanics therefore gives
quite a definite answer to the question of the effect of the
collision, but there is no question of any causal description. One
gets no answer to the question`, `what is the state after the
collision', but only to the question, `how probable is a specified
outcome of the collision' [...]. Here the whole problem of
determinism comes up. From the standpoint of our quantum mechanics
there is no quantity which in any individual case causally fixes the
consequence of the collision; but also experimentally we have so far
no reason to believe that there are some inner properties of the
atom which condition a definite outcome for the collision. [...] I
myself am inclined to give up determinism in the world of the atoms.
But that is a philosophical question for which physical arguments
alone are not decisive.}'' \cite[p. 54]{Born26}
\end{quotation}
Born's interpretation did not seem any less problematic, since it did not account for the interaction of these ``probability waves'' nor the fact that these waves were represented in configuration space and the dimension changed with the number of wave functions that could be considered in different situations. However, more or less in the same way that Bohr had already misdirected the problems of his atomic model, Born's focus switched completely the attention from the intensive patterns considered by Heisenberg to a unilateral attention on the explanation of single outcomes. This, in fact, was due to the prevalence of the atomist presupposition that Bohr had successfully championed: the single outcome was taken, without any justification, as the effect of the presence (or absence) of a particle. And, following this line of reasoning, since we are talking here about a theory that refers to particles, we should concentrate on explaining their natural consequence as single outcomes. This, of course, since the original formalism talked about intensities, presented a consistency problem, an essential difference between what the theory described in formal terms and what it was supposed to deliver in terms of observation. So, in short, the probabilistic interpretation defended by Born connected implicitly the mathematical probability of the theory with the observation of a specific effect of a single particle. Let us add that this reference to elementary particles wan not only completely ungrounded but had no connection whatsoever to the consistent formal-conceptual definition of `particle' that could be found in classical mechanics. In this case, while mathematically a particle was represented in terms of well defined invariant properties ---through Galilean transformations---, in conceptual terms a particle was conceived ---following Aristotle's and Kant's categories--- as an {\it entity} in the actual mode of existence ---namely, as constrained and defined by specific logical and metaphysical principles such as those of non-contradiction and identity. In QM, on the contrary, regardless of the complete lack of a systematic conceptual and mathematical foundation that would support the reference to particles, the atomist narrative was inconsistently ---and dogmatically--- maintained. A way out of trouble has been sometimes to take this reference perhaps not as an ontological account of ``reality-in-itself'', but at least as a ``useful fiction''. But even if we thus play down the importance of the atomist presupposition by taking it as a mere ``tool'', the deductions, methodological and operational steps that are implicit in the atomist representation, continue to function with the same force. As Faraday explained long ago: ``the word \emph{atom}, which can never be used without involving much that is purely hypothetical, is often intended to be used to express a simple fact; but good as the intention is, I have not yet found a mind that did habitually separate it from its accompanying temptations'' \cite[p. 220]{Laudan81}. Even Schrödinger rephrased this idea for the quantum case: ``We have taken over from previous theory the idea of a particle and all the technical language concerning it. This idea is inadequate. It constantly drives our mind to ask information which has obviously no significance'' \cite[p. 188]{Schr50}. 

Before advancing let us note some of the conclusions of this section. Firstly, that Schrödinger’s formulation gained the acceptance of the physics community because of its promise of restoring a somewhat classical {\it continuous} representation. Secondly, that this preferred formulation, however, soon found some definitive obstacles ---as the fact that its differential equation had to be considered within {\it configuration space}--- that Schrödinger himself would recognize. Thirdly, that an equivalence myth between matrix mechanics and Schrödinger’s differential equation was uncritically accepted without a rigorous mathematical justification. And, finally, that the success of Born’s probabilistic interpretation of Schrödinger’s equation, on the one hand, would reinforce the presupposed reference to ``quantum particles'', shifting the focus from intensive patterns (in Heisenberg’s formulation) to single outcomes, and, on the other hand, would find a crucial obstacle when facing the difficulty to define the entirely non-classical ``probability'' that would unfold in this context.

% Let us also stress that the bra-ket notation involves always the implicit reference to a specific basis.

\section{Dirac's Vectorial Axiomatization: Certainty and Contextuality}

An essential aspect of Dirac's 1930's proposed axiomatization of QM is the complete replacement of matrices by vectors in accordance with his {\it transformation theory}. This replacement is already explicit within Dirac's notation where a unit vector in a specific basis is written as a {\it ket} $|x\rangle$. Of course, as it is well known, a {\it ket} can be also seen as a rank one matrix through the following operation $|x\rangle\langle x|$. If $H$ is an $n$-dimensional complex vector space, $H=\mathbb{C}^n$, and $B(H)$ is the space of $n\times n$ matrices, then we can relate the space of vectors $H$ with the space of (rank 1) matrices $B(H)$ through the map:
\[
\nu:H\to B(H),\quad \nu(|x\rangle):=|x\rangle\langle x|.
\]
Let us mention two relevant properties of $\nu$. The first one is that $\nu$ is not {\it surjective}. In fact,  its image is equal to the set of rank one matrices. The second relevant property of $\nu$ is that it is {\it injective}. Hence, we can think of the space of vectors as a subset of the space of matrices. In other words, the vector space is much ``smaller'' than the matrix space (see \cite{deRondeMassri24}). As a matter of fact, this jump from matrices to vectors was already implicit in Schr\"odinger's equation, which could be naturally read as a vectorial equation. Indeed, following Dirac's notation, the  equation governed the time evolution of a vector $|v_t\rangle$ in a multi-dimensional vectorial space reads as follows: 
\[
i \hbar\frac{d}{dt}|v_t\rangle = \hat{H}|v_t\rangle.
\]
Dirac reformulated the mathematical content of the theory exclusively in terms of vectorial spaces. From a mathematical perspective, since vectors can re-generate the whole space of matrices (through the convex sum of rank 1 matrices), there seems to exist a complete equivalence with no loss of information. In fact, one can provide a definition of Hilbert spaces and dimension using sets and functions. However, this {\it mathematical equivalence} has led people to believe that there was also a {\it physical equivalence}, when this is not the case. This becomes clear once we recognize that in Heisenberg's original formulation each and every matrix, of any rank, in each and every basis, possesses a clear operational content which can be directly linked to a particular experimental set up and consequently to a set of definite and observable intensive measurement results (see \cite[Appendix A]{deRondeMassri24}). The elimination of this infinitely many number of matrices implies also the elimination of an enormous amount of experimental data related within the theory to actual experimental situations. Why was that enormous amount of experimental phenomena not significant enough to be considered? 

There has been an unacceptable reduction of the phenomena considered as physically meaningful which, later on, have been re-introduced within SQM in terms of the notion of {\it mixture} (i.e., rank $\neq 1$ matrices) and re-interpreted (in a mere epistemological, non-ontological manner, as not fully real) as statistical convex sums of {\it pure states} (i.e., matrices of rank = 1 or vectors) ---showing the dependency of the whole formulation with respect to vectors represented by single {\it kets} (see \cite[Appendix B]{deRondeMassri24}). The idea that physical situations described by rank $\neq 1$ matrices can be replaced by completely different physical situations described by the convex sum of matrices of rank = 1 imposes an artificial discrimination which is not only alien to the mathematical formalism but is completely unjustified from an experimental perspective  (see \cite{deRondeMassri21b}). In short, a great multiplicity of phenomena, actually observable and captured by the formalism, was simply eliminated to focus only on a very small subset of situations described by {\it pure states} (rank 1 matrices represented by vectors), after which they were reintroduced as ``mixed states'', as if they were in fact the effect of the sum of those pure states previously selected. But why such an interest on reframing everything only in terms of rank 1 matrices or vectors? The reason can be found in Dirac's attempt to follow both Bohr's reference to a microscopic realm in combination with his logical positivist understanding of theories as describing observations. Although Dirac begins with a positivist statement of principles\footnote{Indeed, Dirac  \cite[p. 3]{Dirac74} himself emphasized, already in the first chapter of his famous book, that ``science is concerned only with observable things'' and that ``the main object of physical science is not the provision of pictures, but the formulation of laws governing phenomena and the application of these laws to the discovery of phenomena. If a picture exists, so much the better; but whether a picture exists or not is a matter of only secondary importance.''}, it is evident from the very start of his book that he assumes the Bohrian atomistic representation (as well as his complementarity\footnote{According to the English mathematician: ``quantum mechanics is able to effect a reconciliation of the wave and corpuscular properties of light''  such that ``[a]ll kinds of particles are associated with waves in this way and conversely all wave motion is associated with particles. Thus all particles can be made to exhibit interference effects and all wave motion has its energy in the form of quanta.''}), and the language of particles is present from the beggining. Thus, that supposed intention of adhering to what is observed shifts towards a unilateral emphasis on single, unique, outcomes, as these can be interpreted as the presence (or absence) of those presupposed particles. And it is because rank 1 matrices are those that, when considering the diagonal basis, permit to predict single outcomes with {\it (binary) certainty} ---something completely alien to the matrix formulation which provides an {\it intensive} quantification of phenomena--- that these are taken as the representation of particles, and thus the main element of the formalism. Again, behind this move we find the atomist presupposition made by Bohr and followed by Dirac which allows to switch the focus from the intensive values of matrix mechanics to a unilateral attention on single outcomes, which are then interpreted as the effects of ``particles''. And this operation is done even at the price of an enormous loss of empirical (non-binary) information ---which is later on understood as ``uncertain'' with respect to pure states. Even worse is the fact that this operational restriction to a binary representation is ---unlike the reference to intensities--- dependent on the choice of the {\it reference frame}, and consequently it is not invariant (see \cite{deRondeMassri21a}). As demonstrated by the Kochen-Specker theorem, when we advance with a binary valuation, the possibility of a globally consistent representation becomes precluded and contextuality enters the scene. This goes in line with Dirac's ---also inconsistent--- redefinition of the notion of (quantum) state in terms of a {\it preferred basis} which destroys the possibility of discussing about a state of affairs which remains {\it the same} independently of reference frames. Even worse is the fact that due to the incompatibility between superposed states intensively defined and single measurement outcomes Dirac would impose the existence of ``collapses'', creating a new non-linear evolution triggered by measurement and incompatible with the already existent linear motion of the theory ---encapsulated in Schr\"odinger's wave equation.

\section{Back to Heisenberg and Beyond}

As we have shown, immediately after matrix mechanics, there was first an attempt to return to a spatiotemporal wave representation through the construction of a differential wave equation, followed by a vectorial re-formulation grounded on the atomistic reference to corpuscles, which also made us replace the focus on intensities ---which had allowed Heisenberg to produce an invariant formalism in the first place--- by a unilateral attention to single (bianry) outcomes. The result: an extreme impoverishment of the empirical phenomena captured by the theory, the {\it ad hoc} introduction of a ``collapse'' that would turn the evolution of the theory inconsistent, the contextual redefinition of the notion of (quantum) {\it state} that would preclude the invariant ---basis independent--- reference to a state of affairs, the reintroduction of matrices of rank $\neq$ 1 but now inconsistently understood as ``statistical mixtures'' and the imposition of a microscopic narrative only maintained due to the desire to remain attached to a familiar representation (see for a detailed analysis \cite{deRondeMassri24}). In contradistinction, our proposal is to go against these successive steps which have taken us away from matrix mechanics, to escape the enthronement of {\it pure states} \cite{deRondeMassri21b}, the unilateral focus on single outcomes, the need of unjustified ``collapses'',  the atomist picture, and thus return to the richness of intensive phenomena and the operational-invariance already captured by Heisenberg's matrix formulation itself.

In the history of physics it has been invariance what has always allowed to determine, in formal terms, a \emph{moment of unity} consistent throughout the different {\it reference frames} and experimental situations, thus allowing for an intelligible representation detached from particular, individual perspectives. This has always been seen, by physicists such as Galileo, Newton, Maxwell, and of course by Einstein when developing relativity theory, as an obligation, as a necessary condition in order to construct a consistent physical theory. In fact, Einstein, when developing his special theory of relativity, was faced with the incompatibility among three different conditions if they were maintained simultaneously: between the principle of relativity (the requirement to consistently translate the experiments in one reference frame to another equivalent one), the experimental finding of the constant speed of light, and Galilean transformations \cite{Einstein20}. In order to maintain the principle of relativity (and thus invariance) as well as the experimental evidence of the constant speed of light, Einstein decided to abandon the ``commonsensical'' spatiotemporal classical representation, producing a conceptual innovation completely alien to classical physics. Fidelity to the irreducible conditions of a physical theory seemed rightly more important to him than fidelity to the picture that had been created by classical mechanics just a few centuries before. As a consequence, while in classical mechanics the spatial and temporal values were considered absolute (independent of reference frames) and speed and position as relative (to reference frames), in relativity theory it is the speed of light that would become absolute (independent of reference frames) and spatial and temporal intervals relative (to each reference frame). Of course, it should be stressed that the relative aspects involved in both theories did not imply any inconsistency, since in each case the relative values of properties would be consistently considered in terms of a {\it global transformation}, namely, the Galilean transformation in the case of classical mechanics and the Lorentz transformation in the case of special relativity. And it is this global aspect provided by invariance that would allow in both cases to retain a consistent representation where all reference frames remain of course completely equivalent, this is, related consistently to the same objective state of affairs. In the case of QM, a similar situation appears, and the price to pay to maintain invariance is to remain close to the (intensive) experimental evidence while abandoning the reference to binary outcomes and the classical atomist representation. This entails, as in the case of Einstein, to produce a radical conceptual innovation, namely, the need to develop a new concept of an intensive nature which becomes the natural referent of the theory. This path has already been followed in several papers \cite{deRondeMassri19, deRondeMassri21a, deRondeMassri23}, resulting in the development of a conceptual representation centered around the notion of \emph{power of action}. The power of action is an originally intensive physical concept which cannot be understood as depending on supposedly more fundamental constituents. Departing completely from the classical representation, QM is about the relation between intensive quantities of action. These are the basic element of the theory. And the specific intensive value of each power is given by its \emph{intensity }or \emph{potentia}. In this case, an intensive valuation remains consistent when considered from different reference frames. Furthermore, these intensive physical elements can be perfectly measured in the lab without the need to break the causal evolution of the represented state of affairs. There are no ``quantum jumps'' or ``collapses'' required in order to make sense of measurements which, in this scheme, become once again ---like in the rest of pre-Bohrian-positivist physics--- completely detached from the theoretical representation of the state of affairs. In this case, single outcomes are not the main thing to explain, but are rather understood as just partial, incomplete expressions of intensive values ---and thus require repetition. Like in all physical theories the {\it repetition} of an experimental procedure which keeps referring to {\it the same} state of affairs remains a pre-condition for physical representation itself. Contrary to SQM, the state of affairs represented in terms of a specific set of powers of action with a definite potentia does not change when being measured. As we have shown, this path, which goes back to the invariance of intensities already present in Heisenberg’s formulation and develops it further, permits to restore a global (intensive) valuation, escaping contextuality, and reinstating an invariant physical theory where there is no need for any ``preferred basis'' nor for a ``projection postulate''. 

In any case, in this context we want to advance another argument for the return to matrix mechanics, an argument this time related to the impoverishment of the theory caused by the radical restriction, imposed by the vectorial proposal of Dirac, of the matrices that can be considered as physically meaningful. As we have shown, an enormous amount of empirical information, of phenomena that can be observed, captured and predicted by the theory, was left aside or taken as secondary simply because it did not comply to the expectations of an atomist binary representation. The physical situations captured by the original matrix mechanics are enormously superior than what can be considered in a vectorial formulation that focuses on ``pure states'' ---a notion which is in itself inconsistently defined \cite{deRondeMassri21b, deRondeMassri24}. This, we believe, is an argument in favor of matrix mechanics that cannot be left aside. However, what we propose is not just a return to Heisenberg's matrix mechanics but to produce a natural extension of it to a  tensorial formulation. A formulation where the proposed mathematical formalism is also supplemented with a system of well defined operational physical concepts.

\section{Tensorial Quantum Mechanics}

Following the thread of invariance in QM means to realize that the invariant elements of the theory, the \emph{moments of unity}, are of an intensive nature, that invariance is only destroyed by imposing a completely alien binary valuation, and that it is thus necessary to produce a conceptual representation centered around an intensive notion. For this purpose the concept of {\it power of action} has been proposed, along with other notions, such as the {\it intensity} or {\it potentia} and that of an {\it Intensive State of Affairs}. But equally essential is to develop the notions that allow to precisely relate these concepts to what is experienced, to exactly show how powers of actions can be in fact observed in the lab. As Einstein emphasized, this operational aspect, the necessity to relate the concepts to their corresponding observable phenomena, is an essential precondition when defining meaningful physical concepts. Following this line of reasoning, the concepts of {\it quantum lab, screen, detector} and {\it experimental arrangement} have been introduced in \cite{deRondeFMMassri24a}. Furthermore, there are two important theorems which also allow to secure the invariant relation between bases and factorizations within the matrix formalism \cite{deRondeMassri23}. It is from this standpoint that we are now ready to extend the mathematical formulation of QM beyond matrices. 

An essential point of the tensorial formulation we are now ready to present is its capability not only to take into account the phenomena captured by the matrix formulation but to extend even further the consideration of quantum phenomena in the lab. As we will see in the following, while the situation represented by a vector (or a rank 1 matrix) is limited to an experimental arrangement with only one screen and a matrix to an experimental arrangement with two screens, a tensor is capable to describe the general case of an experimental arrangement with $n$ screens (and multiple detectors). Such a tensorial formulation could allow, for instance ---as we will show in a coming article \cite{deRondeFMMassri24d}---, among other things, for a simple and natural understanding of the phenomena of quantum entanglement when considering multiple screens, something that has remained an open problem within the orthodox literature ---referred to as `multi-partite entanglement'' \cite{xie2024}.

Let us begin with the definition of a (simple) \emph{graph} as a pair $G = (V, E)$, where $V$ is a set whose elements are called vertices (or nodes), and $E$ is a set of unordered pairs $\{v,w\}$ of vertices, whose elements are called edges. While each {\it vertex} is related to the mathematical notion of {\it projector operator} and to the physical concept of  {\it power of action}, each {\it edge} is linked to the mathematical concept of {\it commutation} and the experimental compatibility of powers within a particular measurement set up. 

\begin{definition} {\bf Graph of powers:} Given a Hilbert space $H$, the graph of powers $G(H)$ is defined such that the vertices are the projectors on $H$ (called \emph{powers}), and an edge exists between projectors $P_1$ and $P_2$ if they commute.
\end{definition} 

\noindent It is these powers, in their multiplicity and their relationships, that allow us to define an \emph{Intensive State of Affairs} (ISA) ---that contrasts to the binary {\it Actual State of Affairs} (ASA) that represents situations in classical physics and relativity theory. But first, we need to formalize the notion of {\it intensity} (or {\it potentia}). In general, the assignment of intensities is called Global Intensive Valuation (GIV).

\begin{definition} {\bf Global Intensive Valuation:}  A Global Intensive Valuation is a map from $G(H)$ to the interval $[0,1]$.
\end{definition}

\noindent Clearly, not all GIVs are compatible or consistent with the relations between powers. Thus, we will focus on those that define an ISA as follows:

\begin{definition} {\bf Intensive State of Affairs:} Let $H$ be a Hilbert space of infinite dimension. An \emph{Intensive State of Affairs} is a GIV $\Psi: G(H)\to[0,1]$ from the graph of powers $G(H)$
such that $\Psi(I)=1$ and 
\[
\Psi(\sum_{i=1}^{\infty} P_i)=\sum_{i=1}^\infty \Psi(P_i)
\]
for any piecewise orthogonal operator $\{P_i\}_{i=1}^{\infty}$. The numbers $\Psi(P) \in [0,1]$ are called {\it intensities} or {\it potentia} and the vertices $P$ are called \emph{powers of action}. Taking into consideration the ISAs, it is then possible to advance towards a consistent GIV which can bypass the contextuality expressed by the Kochen-Specker Theorem \cite{deRondeMassri21b, KS}. 
\end{definition} 

\begin{definition} {\bf Quantum Laboratory:} We use the term \emph{quantum laboratory} (or quantum lab or Q-Lab) as the operational concept of an ISA. 
\end{definition} 

\noindent Within a Q-Lab, we have the concepts of {\it screen}, {\it detector} with which we can define more explicitly the notions of {\it power of action}, {\it potentia} and also specify what is an {\it experimental arrangement}: 

\begin{definition}{\bf Screen and Detector:} A \emph{screen} with $n$ places for $n$ detectors corresponds to the vector space $\mathbb{C}^n$. Choosing a basis, say $\{|1\rangle,\dots,|n\rangle\}$, is the same as choosing a specific set of $n$ {\it detectors}. A \emph{factorization} $\mathbb{C}^{i_1}\otimes\dots \otimes\mathbb{C}^{i_n}$ is the specific number $n$ of screens, where the screen number $k$ has $i_k$ places for detectors, $k=1,\dots,n$. Choosing a \emph{basis} in each factor corresponds to choosing the specific detectors; for instance $|\uparrow\rangle, |\downarrow\rangle$. After choosing  a basis in each factor, we get a basis of the factorization $\mathbb{C}^{i_1}\otimes\dots \otimes\mathbb{C}^{i_n}$
that we denote as
\[
\{ |k_1\dots k_n\rangle \}_{1\le k_j\le i_j}.
\]
\end{definition} 

\begin{definition}{\bf Power of action:} The \emph{ basis element} $|k_1\dots k_n\rangle$ determines the \emph{ projector}  $|k_1\dots k_n\rangle \langle k_1\dots k_n|$ which is the formal-invariant counterpart of the objective physical concept called \emph{ power of action} (or simply \emph{power}) that produces a global effect in the $k_1$ detector of the screen $1$,  in the $k_2$ detector of the screen $2$ and so on until the $k_n$ detector of the screen $n$. Let us stress the fact that this effectuation does not allow an explanation in terms of particles within classical space and time. Instead, this is explained as a characteristic feature of powers. In general, any given power will produce an intensive multi-screen non-local effect. 
\end{definition} 

\begin{definition} {\bf Experimental Arrangement:} Given an ISA, $\Psi$, a factorization $\mathbb{C}^{i_1}\otimes\dots \otimes\mathbb{C}^{i_n}$ and a basis $B=\{|k_1\dots k_n\rangle\}$ of cardinality $N=i_1\dots i_n$, we define an \emph{ experimental arrangement} denoted $\EA_{\Psi,B}^{N,i_1\dots i_n}$, as a specific choice of screens with detectors together with the potentia of each power, that is,
\[
\EA_{\Psi,B}^{N,i_1\dots i_n}= \sum_{k_1,k_1'=1}^{i_1}\dots \sum_{k_n,k_n'=1}^{i_n} 
\alpha_{k_1,\dots,k_n}^{k_1',\dots,k_n'}|k_1\dots k_n\rangle\langle k_1'\dots k_n'|.
\]
Where the number $N$ is the cardinal of $B$ and is called the \emph{degree of complexity} (or simply \emph{degree}) of the experimental arrangement. 
\end{definition} 
\begin{definition}{\bf Potentia:} The number that accompanies the power $|k_1\dots k_n\rangle \langle k_1\dots k_n|$ is its \emph{ potentia} (or intensity) and the basis $B$ determines the powers defined by the specific choice of screens and detectors. 
\end{definition} 

Now, assume that in a Q-Lab we want to change or modify an experimental setup. We also have two theorems that allow us to predict the possible outcomes that will be obtained in a new experimental arrangement. If the number of powers (i.e., the complexity) remains the same after the rearrangement, then the {\it Basis Invariance Theorem} tell us that the new experimental arrangement is equivalent to the previous one. However, if the complexity of the new experimental arrangement drops, then the {\it Factorization Invariance Theorem} tell us that all the knowledge in the new experimental arrangement was already contained in the previous one (see \cite{deRondeFMMassri24a}). 

\begin{theorem}{\sc (Basis Invariance Theorem)}
Given a specific QLab $\Psi$, all experimental arrangements of the same complexity, are equivalent independently of the basis. 
\end{theorem}

\begin{theorem} {\sc (Factorization Invariance Theorem)}
The experiments performed within an $\EA_{\Psi}^N$ can also be performed with an experimental arrangement of higher complexity N+M, $\EA_{\Psi}^{N+M}$, that can be produced within the same QLab $\Psi$.  
\end{theorem}

Now, in order to treat this situation formally, we must work with {\it tensors}. Specifically, assume that we have two bases $B$ and $B'$ obtained from two experimental arrangements in the same Q-Lab $\Psi$,
\[
B = \{|k_1\dots k_n\rangle\}_{1\le k_j\le i_j}, \quad
B' = \{|\kappa_1\dots \kappa_{m}\rangle\}_{1\le \kappa_j\le \iota_j}, \quad
\]
From $B$ we infer that the first experimental arrangement has $n$ screens, where the first screen has $i_1$ detectors, the second $i_2$ detectors and so on. The second experimental arrangement has $m$ screens, where the first screen has $\iota_1$ detectors, the second 
$\iota_2$ and so on. Assume that the two bases are related by the following transformation,
\[
|k_1\dots k_n\rangle = \sum_{\kappa_1,\dots,\kappa_m=1} ^{\iota_1,\dots,\iota_m}
\lambda_{\kappa_1,\dots,\kappa_m}^{k_1\dots k_n}
|\kappa_1\dots \kappa_{m}\rangle
,\quad 1\le k_1\le i_1,\dots,1\le k_n\le i_n.
\]
Then, we can convert the first experimental arrangement into the second one through the algebraic properties of the tensors. If
\[
\EA_{\Psi,B} = \sum_{k_1,k_1'=1}^{i_1}\dots \sum_{k_n,k_n'=1}^{i_n} 
\alpha_{k_1,\dots,k_n}^{k_1',\dots,k_n'}|k_1\dots k_n\rangle\langle k_1'\dots k_n'|
\]
then,
\[
\EA_{\Psi,B'} = 
 \sum_{\kappa_1,\dots,\kappa_m=1} ^{\iota_1,\dots,\iota_m}
 \sum_{\kappa_1',\dots,\kappa_m'=1} ^{\iota_1,\dots,\iota_m}
\left(
\sum_{k_1,k_1'=1}^{i_1}\dots \sum_{k_n,k_n'=1}^{i_n} 
\alpha_{k_1,\dots,k_n}^{k_1',\dots,k_n'}
\lambda_{\kappa_1,\dots,\kappa_m}^{k_1\dots k_n}
\overline{\lambda_{\kappa_1',\dots,\kappa_m'}^{k_1'\dots k_n'}}
\right)
|\kappa_1\dots \kappa_{m}\rangle
\langle\kappa_1'\dots \kappa_{m}'|.
\]
Notice that in standard multi-index notation (a notation better suited for these type of algebraic expressions) the last equation can be written more compactly as
\[
\EA_{\Psi,B'} = \sum_{\kappa,\kappa'}
\left(\sum_{k,k'}
\alpha_{k}^{k'}
\lambda_{\kappa}^{k}
\overline{\lambda_{\kappa'}^{k'}}
\right)
|\kappa\rangle\langle\kappa'|.
\]
It now becomes clear that this new formulation produces a bridge between the mathematical formalism and the concepts required in order to refer in a meaningful manner to the experience produced within the laboratory. As we mentioned before, this scheme also opens the doors to the natural consideration of multi-screen entanglement (discussed in the orthodox literature as ``multi-partite entanglement'').

\smallskip 

To sum up, let us summarize all the notions we have discussed showing the relation between the physical and the mathematical representations in the following table:
\begin{center}
\begin{tabular}{|l|l|}
\hline
Physical concepts&Mathematical concepts\\
\hline
Q-Lab $\Psi$ & ISA $\Psi:G(H)\to[0,1]$\\
Screen with $n$ places &   $\mathbb{C}^n$\\
$n$ detectors &    Basis $\{|1\rangle,\dots,|n\rangle\}$\\
$n$ screens where screen $j$ has $i_j$ places &    Factorization  $\mathbb{C}^{i_1}\otimes\dots \otimes\mathbb{C}^{i_n}$  \\
Detectors in each screen &  $ \{ |k_1\dots k_n\rangle \}_{1\le k_j\le i_j}$ \\
Power of action & Projector $|k_1\dots k_n\rangle \langle k_1\dots k_n|$ \\
An experimental arrangement& The tensor $\EA_{\Psi,B}^{N,i_1\dots i_n}$\\
Degree of complexity $N$ & Cardinal of $B$\\
Potentia of the power  &
The real number $\alpha_{k_1,\dots,k_n}^{k_1,\dots,k_n}$\\
Modify an experimental arrangement&Change of basis\\
\hline
\end{tabular}
\captionof{table}{Relations between physical and mathematical concepts.}
\end{center}

\section{Final Remarks} 

As we have shown, the replacement of matrix mechanics by a vectorial formulation implied the loss of an enormous amount of  physically meaningful information about experimental situations and phenomena. We presented this impoverishment of the theory as an argument for a return to the matrix formalism. But, as we have shown in the last section, if we advance even further into a tensorial formulation, we can not only recover the phenomena captured by matrix mechanics, but also extend its reach and consider even more experimental situations that can be observed in the lab. Notice that while the case of a single screen corresponds to the orthodox vectorial approach, the case of two screens is linked to the orthodox extension to density matrices. It is important to remark that in our tensional formulation these are particular cases of the more general situation where we have $n$ screens. We are thus able, within Tensorial QM, to consider as many screens as we want without any complications. This leads, for example, to the possibility to naturally consider a multi-screen analysis of entanglement which is capable to evade the many problems found when attempting to measure quantum entanglement \cite{deRondeFMMassri24b} and also within the multi-partite orthodox account of entanglement \cite{deRondeFMMassri24d}. To conclude, we can say that Tensorial QM enables us not only to regain the phenomena captured by matrix mechanics, but also to extend our grasp in order to consider a wider amount oh phenomena, a greater number of experimental situations that can be thus incorporated without any difficulty. Finally, the rigorous conceptual development of a specific system of notions capable to account for the mathematical formalism in an operational fashion opens the door to an intuitive understanding of the theory of quanta.

\section*{Acknowledgements} 

This work was partially supported by the following grant: ANID-FONDECYT, project number: 3240436. The authors state that there is no conflict of interest.


\begin{thebibliography}{1}

\bibitem{Bohr48} Bohr, N., 1948, ``On the notions of causality and complementarity'', {\it Dialectica}, {\bf 2}, 312-319.

\bibitem{Bohr60} Bohr, N., 1960, {\it The Unity of Human Knowledge}, In {\it Philosophical writings of Neils Bohr}, vol. 3., Ox Bow Press, Woodbridge.

\bibitem{Bohr87} Bohr, N., 1987, {\it The Philosophical Writings of Niels Bohr}, 3 Volumes, Ox Bow Press, Woodbridge.

\bibitem{Bokulich06} Bokulich, A., 2006, ``Heisenberg Meets Kuhn: Closed Theories and Paradigms'', {\it Philosophy of Science}, {\bf 73}, 90-107.

\bibitem{BokulichBokulich20} Bokulich, A. $\&$ Bokulich, P., 2020, ``Bohr's Correspondence Principle'', {\it The Stanford Encyclopedia of Philosophy (Fall 2020 Edition)}, E.N. Zalta (ed.). https://plato.stanford.edu/archives/fall2020/entries/bohr-correspondence/.

\bibitem{Born26} Born, M., 1926, ``Zur Quantemechanik der Stossvorgange'', {\it Zeitschrift f\"ur Physik} {\bf 37} (1926) 863-867, English translation ``On the Quantum Mechanics of Collision Processes'' in \cite{WZ}.

%\bibitem{Born71} Born, M., 1971, {\it The Born-Einstein Letters}, Walker and Company, New York.

\bibitem{BornJordan25} Born, M. $\&$ Jordan, P., 1925, ``Zur Quantenmechanik'', Zeitschrift f\"{u}r Physik \textbf{34}, 858-888. English translation in \emph{Sources of Quantum Mechanics}, B.L. van der Waerden, ed., Dover, 1968.

\bibitem{BornHeisenbergJordan26} Born, M., Heisenberg, W. $\&$ Jordan, P.,1926, ``Zur Quantenmechanik II'', {\it Zeitschrift f\"{u}r Physik}, \textbf{35}, 557-615. English translation in \emph{Sources of Quantum Mechanics}, B.L. van der Waerden, ed., Dover, 1968.

\bibitem{Bub97} Bub, J., 1997, {\it Interpreting the Quantum World}, Cambridge University Press, Cambridge.

\bibitem{deRonde17} de Ronde, C., 2017, ``Causality and the Modeling of the Measurement Process in Quantum Theory'', {\it Disputatio}, {\bf IX}, 657-690.

\bibitem{deRonde18} de Ronde, C., 2018, ``Quantum Superpositions and the Representation of Physical Reality Beyond Measurement Outcomes and Mathematical Structures'', {\it Foundations of Science}, {\bf 23}, 621-648.

\bibitem{deRonde23a} de Ronde, C., 2023, ``Measuring Quantum Superpositions (Or, ``It is only the theory which decides what can be observed.'')'' in {\it Non-reflexive logics, non-individuals, and the philosophy of quantum mechanics: Essays in honour of the philosophy of D\'ecio Krause}, J.R. Arenhart, R. Arroyo (Eds.), Springer, Berlin.

\bibitem{deRonde23b} de Ronde, C., 2023, ``Mythical Thought in Bohr's Anti-Realist Realism (Or: Lessons on How to Capture\\ and Defeat Smoky Dragons)'' in {\it From Contradiction to Defectiveness to Pluralism in Science: Philosophical and Formal Analyses}, O. Bueno and M. Martinez-Ordaz (Eds.), Springer, in press.

\bibitem{deRondeFMMassri24a} de Ronde, C., Fern\'andez Mouj\'an, R. $\&$ Massri, C., 2024, ``Equivalence Relations in Quantum Theory: An Objective Account of Bases and Factorizations'', sent (quant-ph:2404.14891).

\bibitem{deRondeFMMassri24b} de Ronde, C., Fern\'andez Mouj\'an, R. $\&$ Massri, C., 2024, ``Everything is Entangled in Quantum Mechanics: On the Measures of Quantum Entanglement'', sent (quant-ph:2405.05756).

\bibitem{deRondeFMMassri24d} de Ronde, C., Fern\'andez Mouj\'an, R. $\&$ Massri, C., 2024, ``Multiscreen Entanglement in Tensorial Quantum Mechanics'', preprint.

\bibitem{deRondeMassri19} de Ronde, C. $\&$ Massri, C., 2019, ``The Logos Categorical Approach to Quantum Mechanics: II. Quantum Superpositions and Measurement Outcomes.'', {\it International Journal of Theoretical Physics}, {\bf 58}, 1968-1988. 


\bibitem{deRondeMassri21a} de Ronde, C. $\&$ Massri, C., 2021, ``The Logos Categorical Approach to Quantum Mechanics: I. Kochen-Specker Contextuality and Global Intensive Valuations.'', {\it International Journal of Theoretical Physics}, {\bf 60}, 429-456. 

%\bibitem{deRondeFM21} de Ronde, C. $\&$ Fern\'andez-Mouj\'an, R., 2021, ``Are `Particles' in Quantum Mechanics ``Just a Way of Talking''?'', preprint (http://philsci-archive.pitt.edu/19968/). 

\bibitem{deRondeMassri21b} de Ronde, C. $\&$ Massri, C., 2021, ``Against the Tyranny of Pure States in Quantum Theory'', {\it Foundations of Science}, {\bf 27}, 27-41.

\bibitem{deRondeMassri24} de Ronde, C. $\&$ Massri, C., 2022, ``The Many Inconsistencies of the Purity-Mixture Distinction in Standard Quantum Mechanics'', sent (quant-ph:2208.10574).  

\bibitem{deRondeMassri23} de Ronde, C. $\&$ Massri, C., 2023, ``Relational quantum entanglement beyond non-separable and contextual relativism'', {\it Studies in History and Philosophy of Science}, {\bf 97}, 68-78.

\bibitem{Dirac74} Dirac, P.A.M., 1974, {\it The Principles of Quantum Mechanics}, 4th Edition, Oxford University Press, London.

\bibitem{Einstein20} Einstein, A., 1920, {\it Relativity. The Special and General Theory}, Henry Holt $\&$ Company, New York. 

\bibitem{Faye21} Faye, J $\&$ Jaksland R., 2021, ``What Bohr wanted Carnap to learn from quantum mechanics'', {\it Studies in History and Philosophy of Science} {\bf 88}, 110-119.

\bibitem{Feynman67} Feynman, R.P., 1967, {\it The Character of Physical Law}, Massachusetts Institute of Technology Press, Massachusetts. 

\bibitem{Heis25} Heisenberg, W., 1925, ``\"{U}ber Quantentheoretische Umdeutung Kinematischer und Mechanischer Beziehungen'', Zeitschrift f\"{u}r Physik \textbf{33}, 879-893. English translation in \emph{Sources of Quantum Mechanics}, B.L. van der Waerden, ed., Dover, 1968.

%\bibitem{Heis27} Heisenberg, W., 1927, ``Uber den anschaulichen Inhalt der quantentheoretischen Kinematik und Mechanic", {\it Zeitschrift f\"ur Physik}, {\bf 43}, 172-98; reprinted as ``The Physical Content of Quantum Kinematics and Mechanics, in {\it Quantum Theory and Measurement}, J.A. Wheeler and W.H. Zurek (Eds.).

\bibitem{Heis58} Heisenberg, W., 1958, {\it Physics and Philosophy}, George Allen \& Unwin Ltd., London.

\bibitem{Heis58a} Heisenberg, W., 1958,  ``The Representation of Nature in Contemporary Physics'', {\it Daedalus}, {\bf 87}, 95-108.

\bibitem{Heis71} Heisenberg, W., 1971, {\it Physics and Beyond}, Harper \& Row, New York.

%\bibitem{Heis73} Heisenberg, W., 1973, ``Development of Concepts in the History of Quantum Theory'', in {\it The Physicist's Conception of Nature}, pp. 264-275, J. Mehra (Ed.), Reidel, Dordrecht.

\bibitem{HilgevoordUffink01} Hilgevoord, J. $\&$ Uffink, J., 2001, ``The Uncertainty Principle'', {\it The Stanford Encyclopedia of Philosophy (Winter 2001 Edition)}, E. N. Zalta (Ed.), http://plato.stanford.edu/archives/win2001/entries/qt-uncertainty/.

%\bibitem{Howard10} Howard, D, 2010, ``Einstein's Philosophy of Science'', {\it The Stanford Encyclopedia of Philosophy (Summer 2010 Edition)}, E. N. Zalta (Ed.), http://plato.stanford.edu/archives/sum2010/entries/einstein-philscience/.

\bibitem{Jammer89} Jammer, M., 1989, \emph{The conceptual development of quantum mechanics}, Tomash Publishers, New York.

\bibitem{Jones08} Jones, S., 2008, {\it The Quantum Ten}, Oxford University Press, New York.

\bibitem{KS} Kochen, S. $\&$ Specker, E., 1967, ``On the problem of Hidden Variables in Quantum Mechanics'', {\it Journal of Mathematics and Mechanics}, {\bf 17}, 59-87. Reprinted in Hooker, 1975, 293-328.

\bibitem{Laudan81} Laudan, L., 1981, ``Ernst Mach's Opposition to Atomism'', in {\it Science and Hypothesis}, Springer, Dordrecht.

\bibitem{Moore89} Moore, W., 1989, {\it Schr\"0dinger}, Cambridge University Press, New York.

\bibitem{Muller97} Muller, F., 1997, ``The Equivalence Myth of Quantum Mechanics ---Part I'', {\it Studies in History and Philosophy of Modern Physics}, \textbf{28}, 35-61.

\bibitem{Pringe14} Pringe, H., 2014, ``Cassirer and Bohr on Intuitive and Symbolic Knowledge in Quantum Physics'', {\it Theoria}, {\bf 29}, 417-429.

\bibitem{Schlosshauer11} Schlosshauer, M. (Ed.), 2011, {\it Elegance and Enigma. The Quantum Interviews}, Springer-Verlag, Berlin.

\bibitem{Schr28} Schr\"{o}dinger, E., 1928, \emph{Collected papers on wave mechanics}, Blackie and Sons Limited, London.

%\bibitem{Sch1} Schr\"{o}dinger, E., 1926, ``Quantisation as a problem of proper values. Part I'', {\it Annalen der Physik}, \textbf{79}, 361.

%\bibitem{Sch2} Schr\"{o}dinger, E., 1926, ``Quantisation as a problem of proper values. Part II'', {\it Annalen der Physik}, \textbf{79}, 489.

%\bibitem{Sch3} Schr\"{o}dinger, E., 1926, ``Quantisation as a problem of proper values. Part III'', {\it Annalen der Physik}, \textbf{80}, 437.

%\bibitem{Sch4} Schr\"{o}dinger, E., 1926, ``Quantisation as a problem of proper values. Part IV'', {\it Annalen der Physik}, \textbf{81}.

\bibitem{Sch5} Schr\"{o}dinger, E., 1926, ``On the relation between the quantum mechanics of Heisenberg, Born and Jordan, and that of Schr\"{o}dinger'', {\it Annalen der Physik} \textbf{79}, 734. 

\bibitem{schrgas} Schr\"{o}dinger, E., 1926, ``Zur Einsteins Gastheorie'', {\it Physicalische Zeitschrift}, \textbf{27}, 95-101.

\bibitem{Schr35} Schr\"odinger, E., 1935, ``The Present Situation in Quantum Mechanics'', {\it Naturwiss}, {\bf 23}, 807. Translated to english in {\it Quantum Theory and Measurement}, J. A. Wheeler and W. H. Zurek (Eds.), 1983, Princeton University Press, Princeton.

\bibitem{Schr50} Schr\"odinger, E., 1950, ``What is an elementary particle?'', {\it Endeavor}, {\bf 9}, 109-116.

\bibitem{Schr52} Schr\"odinger, E., 1952, ``Are There Quantum Jumps? Part I'', {\it The British Journal for the Philosophy of Science}, {\bf  3}, 109-123.

\bibitem{tomonaga} Tomonaga, S.-I., 1962, \emph{Quantum Mechanics, vol 1}, North Holland Publishing Company, Amsterdam.

%\bibitem{VN} Von Neumann, J., 1996, {\it Mathematical Foundations of Quantum Mechanics}, Princeton University Press (12th. edition), Princeton.

\bibitem{WZ} Wheeler, J. A. $\&$ Zurek, W. H. (Eds.) 1983, {\it Theory and Measurement}, Princeton University Press, Princeton.

\bibitem{xie2024} Xie, S., Younis, D., Mei, Y. $\&$ Eberly, J.H., 2024, ``Multipartite Entanglement: A Journey through Geometry'', {\it Entropy}, {\bf 26}, 217-.

\end{thebibliography}
\end{document}